# A RESILIENT ICT4D APPROACH TO ECO COUNTRIES' EDUCATION RESPONSE DURING COVID-19 PANDEMIC

Azadeh Akbari, Political Geography Working Group, University of Münster, a.akbari@uni-muenster.de

**Abstract:** According to the United Nations, schools' closures have impacted up to 99 per cent of the student population in low and lower-middle-income countries. This research-in-progress report introduces a project on Emergency Remote Teaching (ERT) measures in the ten member states of the Economic Cooperation Organization (ECO) with a focus on the application of Information and Communication Technologies (ICTs) in primary and secondary education levels. The project takes a comparative approach within a resilient ICT-for-Development (ICT4D) framework, where the coping, endurance, and return to pre-crisis functionalities in education systems are studied. The preliminary research demonstrates the impacts of the country's general COVID-19 strategy, the education system in place, and digital infrastructure's level of development on instigating distance-learning platforms. The paper further shows that in addition to access to stable internet connections and digital devices, other infrastructural factors such as access to food, electricity and health services play a significant role in education response planning and implementation. Human factors in the education system, such as teacher training for the usage of ICTs, digital literacy of students and parents, and already existing vulnerabilities in the education system pose challenges to crisis management in the education sector. Other socio-political factors such as attitudes towards girls' education, level of corruption, institutional capacity, and international sanctions or available funds also make the education system less resilient.

## 1. INTRODUCTION

The COVID-19 pandemic has caused the largest disruption of education systems in a historically unprecedented way, "affecting nearly 1.6 billion learners in more than 190 countries and all continents" (United Nations, 2020). Closures of schools have impacted up to 99 per cent of the student population in lower- and lower-middle-income countries (ibid.). Many countries have adopted emergency remote teaching (ERT) methods utilising information and communication technologies (ICTs), including educational platforms, radio and TV lessons, and social media channels. This research-in-progress report introduces a project on ERT measures in the member states of ECO with a focus on the application of ICTs in primary and secondary education levels. Economic Cooperation Organization (ECO) is an intergovernmental regional organization aiming for the sustainable economic development of its ten member states and the region. In addition to shared cultural and historical affinities, the member states use infrastructural and business links to achieve common goals. In March 2017, during ECO's 13th Summit in Islamabad, "ECO Vision 2025" was endorsed (ECO, n.d.). One of the expected outcomes under ECO Vision 2025 in the area of social welfare and environment is the fulfilment of regional mechanisms/frameworks "to support education, training, vocational needs and other capacity enhancement requirements of the peoples/entities of the Member Countries" (ECO, 2017).

Despite such plans and the socio-cultural similarities, the crisis response in the education sector in these countries was not shaped through regional cooperation. It was extremely dependent on the country's general COVID-19 strategy, the education system in place, and digital infrastructure's level of development. The preliminary research of this paper shows that in addition to access to stable internet connections and digital devices, other infrastructural factors such as access to food, electricity and health services play a major role in education planning during the crisis. Human factors in the education system, such as teacher training for the usage of ICTs, digital literacy of students and parents, and already existing vulnerabilities in the education system pose challenges to crisis management in the education sector. Other socio-political factors such as attitudes towards girls' education, level of corruption, institutional capacity, and international sanctions or available funds also complicate the design and implementation of ERT measures.

The research project studies the education and distance learning programmes of each ECO country during the COVID-19 pandemic with a comparative approach within a resilient ICT-for-Development (ICT4D) framework (Heeks & Ospina, 2019). Resilience is understood "as the ability of systems to cope with external shocks and trends" (ibid. p.71) and as a quality that enables not only survival but also endurance throughout the crisis and a return to planned functionalities before the crisis time. Consequently, the research scrutinises resilience attributes such as robustness, self-organisation, learning, redundancy, rapidity, scale, diversity and flexibility, and equality in education crisis responses of ECO countries and investigates the role of ICTs in building resilience through ERT programmes. In the following sections, the report introduces some of the preliminary research on each ECO country's education response but does not provide concrete outcomes as the research is still underway as this paper is being written, The preliminary data is gathered mainly through desk research through media channels, teachers' forums, policy plans, governmental and non-governmental reports, project reports, and especially through evaluations, funding and policy implementation reports of international organisations such as UNICEF, UNESCO, and Global Partnership for Education.

The research project has conducted a saturation review (Rennison & Hart, 2018, p.73) of the English-language literature on distance learning during COVID-19 pandemic until no relevant literature could be identified. From 112 publications (papers, conference submissions, chapters, etc.) written on the subject, 59 were directly addressing issues of distance learning in higher education. Within the remaining papers, 20 were dedicated to the educational challenges during the crisis in developed countries. Others focused on other aspects of education such as teachers training, student engagement, general challenges of e-learning, food insecurity as a result of school closure, the opportunities and shortcoming of distance learning for children with learning disabilities, teachers' attitude towards e-learning, leadership in education during a crisis, using social media and radio for teaching, innovative assessment techniques, lessons learned by different stakeholders, the gap between private and public schools, learners' social interaction, effective knowledge transfer, and different teaching-learning modalities with a focus on blended learning.

The remaining Publications, dedicated to ERT and online distance education (ODE) in primary and secondary educational levels in developing countries during the COVID-19 pandemic researched the challenges, shortcomings, and chances provided by these programmes in Palestine, Libya, and Afghanistan (Khlaif & Salha, 2020), Moldova (Tatiana, 2020), Latin America and the Caribbean (Jaramillo, 2020), China (Zhang, et al., 2021), Nigeria (Chiemeke & Imafidor, 2020), Indonesia (Hidayat, Anisti, Purwadhi, & Wibawa, 2020), Lebanon (Rouadi & FaysalAnouti, 2020), Kenya (Mabeya, 2020), Sri Lanka (Ranasinghe, Kumarasinghe, Somasiri, Wehella, & Kathriarachchi, 2020), African countries (Aseey, 2020), and India (Raj & Khare, 2020) (Bhaumik, 2020). The outcomes of these studies is summarised in the below table.

| | | |
|---|---|---|
| Infrastructural challenges (hard) | fundamental and ICT infrastructure | access to electricity |
| | | access to internet |
| | | access to digital devices |
| | | costs of access, maintenance, and design |
| | | digital learning tools and platforms |
| | housing and access to shelter | lack of private space |
| | | stable location for nomadic communities |
| Organisational challenges | | lack of national crisis response plan for education sector |
| | | funding problems |
| | | cooperation among different stakeholders |
| | | teacher training programmes |
| human development challenges (soft) | teaching capacities | quality of teaching content |
| | | evaluation and assessment |
| | | encouraging student participation |
| | | learning resources |
| | | teaching approaches |
| | | learning strategies |
| | | supports and services for teachers and students |
| | students' capacity | learning adaptability |
| | | participation and motivation |
| | | lack of, or insufficient, parental supervision |
| | digital literacy | teachers |
| | | parents |
| | | students |
| increased exposure to harm | | digital privacy and data protection |
| | | domestic violence |
| | | girls' pregnancy |
| | | mental health |
| | | food insecurity as a result of school closure |

**Table 1. Findings' summary: publications on ERT in developing countries' schools during the COVID-19 pandemic**

The preliminary research in ECO countries shows similar challenges although it discovers some unexplored aspects such as challenges and chances for girls' education in the region as a result of ODE programmes. The attitude towards girls' education and the state of girls' school attendance in ECO countries underlines the importance of taking a comparative approach to countries with socio-cultural similarities. Other collective projects have also taken a comparative approach amongst a variety of countries regardless of development or income-level. For example, in a comparative study between 31 countries (Bozkurt, et al., 2020) general areas of concerns are identified as psychological pressure, emerging educational roles of the parents, support communities and mechanisms, pedagogy of care, alternative assessment and evaluation method, data privacy concerns, digital divide, inequity and social justice, open educational practices and resources, gender issues, and essential (soft) skills and competencies to survive in a time of crisis. Although almost every country in the world has taken ERT measures to sustain educational and schooling programmes, the crisis response is fundamentally different between resourceful and lower- and lower-middle income countries. The regional choice in this research project, therefore, aims to overcome such fundamental disparities.

## 2. School Education During COVID-19 Pandemic

In this section, ECO member states' response to the COVID-19 pandemic regarding school education will be discussed in more detail. Each country profile provides a short overview of the education statistics and main challenges in the country and the plans and programmes implemented as part of the national crisis management. As mentioned earlier, the research project is not yet concluded and the descriptive details will be evaluated in a resilient ICT4D framework to assess the resilience of school systems in ECO region with a focus on ICT tools and programmes.

### 2.1. Afghanistan

The number of children in school in Afghanistan has increased by almost nine times since 2001, and has grown to more than 9.2 million students in 2015, of which 39% are girls (Ministry of Education of the Islamic Republic of Afghanistan, 2016). Although the number of schools has increased from 3,400 to 16,400 (ibid.), 3.7 million children in Afghanistan, or nearly half of all school-age children, are not in formal education (Human Rights Watch, 2020). The nation-wide school closures in Afghanistan has exacerbated an already vulnerable education system. Even though Afghanistan has received more than $100 million in global aid to alleviate the impacts of COVID-19, private educational institutions receive no support from the government (Dawi, 2020). With schools closed for months, vulnerable children such as girls and child labourers are exposed to a higher risk of dropping out. UN Women, UNICEF and Human Rights Watch jointly issued an alert to highlight the gender-specific impact of COVID-19 in Afghanistan (Human Rights Watch, 2020), where Approximately 60% of out-of-school children are girls (UNICEF Afghanistan, n.d.).

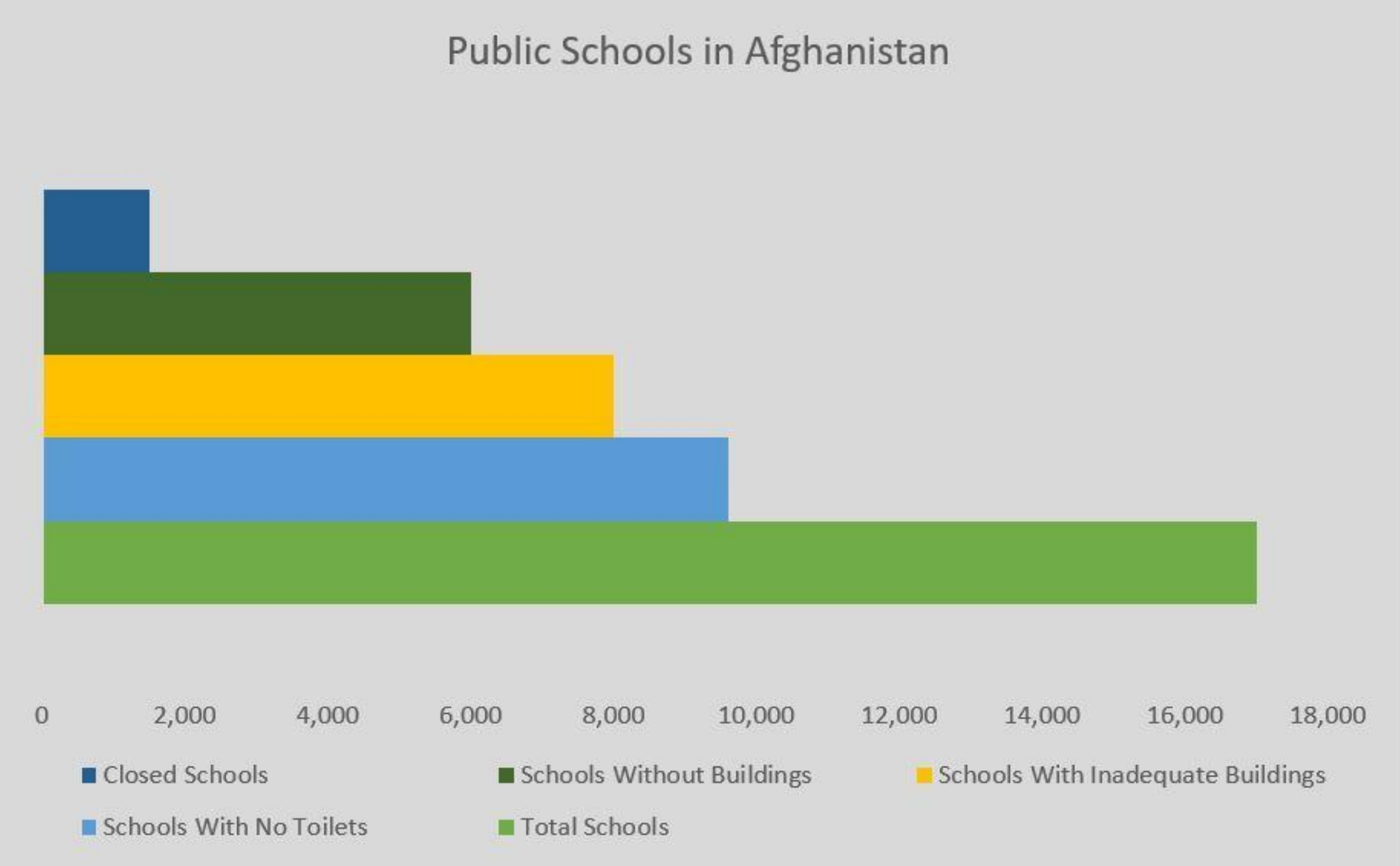

**Figure 1. School Facilities in Afghanistan based on data from the Afghan Education Ministry, Human Rights Watch, PenPath, and the World Bank (Mehrdad, 2020).**

Meanwhile, the education ministry's efforts to promote distance learning using radio and TV stations remain futile since almost 70 per cent of the population has no electricity access (Mehrdad, 2020). The lack of electricity is one of the many infrastructural shortcomings in the education sector, such as sanitation facilities, roads and transportation, school buildings, etc (Figure 1). Institutional capacities such as the insufficient number of teachers, especially female teachers, low quality of teaching, and shortage of textbooks also make an emergency response to a pandemic challenging.

Other cultural factors such as girls' early marriage or patriarchal beliefs against sending girls to school become increasingly difficult to tackle during a health crisis.

### 2.2. Azerbaijan

The schools are reopening gradually in Azerbaijan after a shutdown since March 2020 and a nation-wide distance learning programme. In this phase students will attend school three times a week and on the remaining days teaching will be online (JAMnews, 2021). During the lock-down national TV channels broadcasted TV classes as well as other creative arts competitions and programmes. Recorded TV programs were also available online on a learning platform of the Ministry of Education (United Nations Human Rights, 2020). The country also launched a National Platform called Virtual School in partnership with Microsoft in April 2020, but the enrollment was voluntary and relied upon each school's decision (azərbaycan müəllimi, 2020). Local companies BestComp, AzEduNet, and Edumedia were involved in addressing possible technical issues (ibid.). In addition to training videos, the Ministry of Education provided ICT-training for 14,000 teachers and Azercell, a local mobile network operator, agreed with the Ministry of Transport, Communications and High Technologies to provide a free internet package to 30,000 teachers (ibid.). Azerbaijan's cooperation with international institutions such as UNICEF helped the timely distribution of support material for children, particularly those with special needs, teachers, and parents (UNICEF Azerbaijan, n.d.). With a school enrollment ratio of almost 90% (UNESCO, 2021) and 80% of households enjoying Internet access (The State Statistical Committee of the Republic of Azerbaijan, 2020), the education system was resilient in response to the COVID-19 pandemic.

### 2.3. Iran

Iran launched its national virtual school platform at the end of September 2020, but according to official statistics, 38 per cent of students are not connected to it, and an additional 20 per cent have registered but are not active on the platform (Radio Zamaneh 2020). 21 per cent of students have no means of acquiring a digital device to access the platform (ibid.). Besides the shortcomings in digital devices, accessing internet is also challenging. In many villages, students have to climb nearby mountains to receive internet coverage on their mobile phones (Yousefi, 2020). Iran has supported privatisation of the education sector in the last decades, and currently, 12% of students study at private schools (Mehr News, 2020). These schools use more efficient technological platforms such as Webynar[1], which offer local techno-solutions in the face of international sanctions that make access to global platforms such as Zoom or Skype impossible (S., 2020). The discrepancy between the resources, infrastructure and quality teaching available to public and private school students widens the educational gap between different economic groups.

### 2.4. Kazakhstan

Kazakhstan started distance-learning programmes concurrent with the first lock-down, and in the first semester of the 2020 school year (from September 1 through November 4), 2.4 million out of 3.4 million students attended virtual classrooms (RFE/RL's Kazakh Service, 2020). With 70 per cent of schoolchildren receiving an online education, the educational platforms have shown progress: BilimLand Online Mektep (Knowledge Land Online School) provides access to complete school programs and libraries (ibid.), but the hybrid system of education with present teaching for children with no access to digital devices, in addition to postal services, radio and TV (UNESCO Almaty, 2020) has put an extra burden on teachers' shoulders. The government has distributed 500,000 computers or laptops to students in low-income families and has pledged to provide 300,000 more (ibid.). Other civil society initiatives such as Connect-Ed donate used or new devices to schoolchildren in need (Gabdulhakov, 2020). The economic disparity between different regions of Kazakhstan contributed to poor education even before the COVID-19 crisis. The World Bank,

[1] https://webynar.ir/

UNICEF, UNESCO[2], and WHO[3], jointly with the Ministry of Education and Science of the Republic of Kazakhstan, focus now on post-COVID learning recovery to hinder the educational disparity worsened after a long phase of unequal access to schooling (Jetpissova, 2020).

### 2.5. Kyrgyz Republic

In December 2020, the UNICEF office in Kyrgyzstan published an appeal for humanitarian action for children (UNICEF, 2020). The appeal highlighted the profound impact of the pandemic on the well-being of children and warned about the "disruptions to education, water, sanitation and hygiene (WASH), nutrition and health – including vaccination – services, and the rise in violence, poverty and stress" (ibid). UNICEF also reported that school closures have affected all 1.6 million school children, which the majority of them will continue remote learning in the 2020/21 school year due to lack of adequate WASH facilities and safety measures in schools (ibid.). Since April, the Ministry of Education and Science, in collaboration with UNICEF and other development partners, has developed a remote learning platform for all students and preschool children (Zhusupova, 2020). More than 200 teachers from Bishkek and Osh prepared video lessons and conducted their classes according to the school curriculum via various TV channels (ibid.). Teachers used other international platforms such as Google Classroom and zoom to provide one-to-one lessons and feedback (Abdieva, 2020), and 75,000 teachers were trained on delivering digital, distance and blended learning (UNICEF, 2020).

### 2.6. Pakistan

All schools were closed in Pakistan as part of a sporadic lock-down from March 2020 until January 2021 (Ejaz, Khaliq, & Bajwa, 2021). The Ministry of Federal Education and Professional Training launched the TeleSchool initiative at the beginning of lock-down in collaboration with local education technology providers such as Knowledge Platform, Sabaq.pk, Sabaq Muse, and Taleemabad to broadcast free learning content to grades 1-12 students (ibid). In December 2020, the government introduced its first radio-school to expand student outreach (ibid.). Although experts warn about the learning losses in Pakistan due to school closures, digital infrastructure disparities, low-quality schooling systems in rural areas and general economic hardships (Geven & Hasan, 2020), others have pointed out the opportunities arisen in distance learning to increase girls' education. In Pakistan, one out of three girls have never been to school as a result of distance from schools, security, and lower numbers of female teachers (Ejaz, Khaliq, & Bajwa, 2021). The demands for distance learning have also contributed to an 80%-growth in education technology start-ups (ibid.) The EdTech Pakistan 2.0 workshop organised by the World Bank used the momentum to connect start-ups with development partners, government, and other stakeholders to focus on distance education, blended learning, and girls' education during and after the pandemic (ibid.).

### 2.7. Tajikistan

Tajikistan denied the arrival of the COVID-19 Pandemic in the country for quite a long time (Putz, 2020a) but took cautionary measures to finish the 2020 school year early and start schools ahead of the traditional September opening in August (Putz, 2020b). At schools, students maintain physical distancing in the schoolyard, wear a mask, when possible organise classes outdoors, and regularly wash their hands (Putz, 2020b). However, the UNICEF office in Tajikistan has received a Global Partnership for Education (GPE) grant to support the Ministry of Education and Science with a National Education Preparedness Response Plan (GPE, 2021). The plan includes additional education interventions, such as developing TV lessons and a distance learning platform (ibid.). The blended learning platform uses LearnIn, Learning Passport, and Maktab Mobile to make the video lessons available online (Ammon, 2020).

[2] The United Nations Educational, Scientific and Cultural Organization

[3] The World Health Organization

## 2.8. Turkey

The Ministry of Education in Turkey developed a digital education platform (EBA) at the early stages of the pandemic and also collaborated with the Turkish Radio and Television Corporation (TRT) to broadcast educational material (Özer, 2020). As an exception amongst other ECO member states, Turkey had already developed the EBA digital educational platform since 2011. The platform offers "various learning materials, including curriculum-based videos, documents, e-books, tests, [and] activities" for students from preschool to high school level (Özer, 2020, p. 1126). The platform also includes assessment tools and online broadcasting features. Students with no internet connection can watch the videos on national TV; "TV programs are transmitted via three different channels for primary, early secondary and secondary school students" (ibid.). The level of crisis preparedness is also evident in the 2018 Teaching and Learning International Survey (TALIS) among OECD (the Organisation for Economic Co-operation and Development) countries. As demonstrated in Figure 2, teachers show a good level of confidence in working with digital platforms. Other supporting measures such as free internet for student or training 125,000 teachers on distance teaching in collaboration with UNESCO (Özer, 2020, p. 1127) maintain a more inclusive crisis response.

Turkey is also home to more than 1.6 million child refugees, out of which 400,000 remained out of formal education even before the COVID-19 pandemic (UNICEF, 2019). UNICEF has established 170 physical EBA Support Centres, including six Mobile Centres in six provinces with a high refugee population and has distributed 90,548 learn-at-home kits (UNICEF Turkey, 2021), but the educational losses experienced as a result of the pandemic, in addition to the existing challenges of the refugee crisis will unfold in the mid and long term.

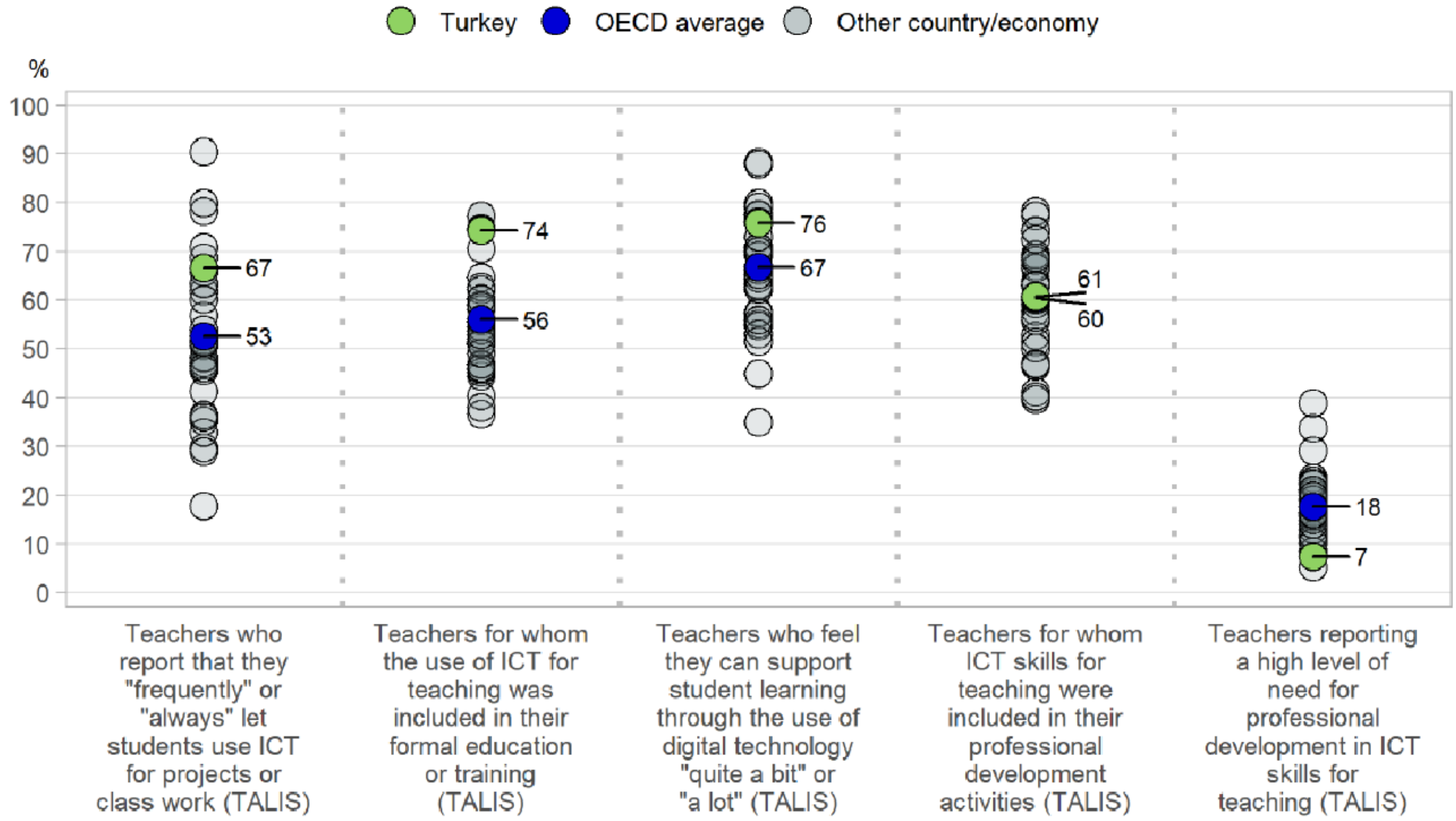

**Figure 2. School and student preparedness for ICT-based learning prior to the crisis- Survey conducted in 2018 (OECD, 2020)**

## 2.9. Turkmenistan

Turkmenistan might be the only country in the world that still denies the arrival of the COVID-19 pandemic within its borders. The government neither implemented a general lock-down nor closed schools or educational facilities. The public measures remained limited to closing the borders, restricting cross-country travel, introducing social distancing, raising awareness on personal hygiene, and limiting access to certain public spaces (Chronicles of Turkmenistan, 2020). In 2017, Turkmenistan adopted the Concept for the Development of a Digital Education System, which aims to improve the quality of educational services through a digital platform and create conditions for continuous education for all segments of the population (Government of Turkmenistan, 2019). Despite such goals for sustainable development, a shaky digital infrastructure and lack of digitall literacy have left this vision not implementable during the pandemic (Muhamedov, 2020).

### 2.10. Uzbekistan

Uzbekistan closed down all schools during its first response to the COVID-19 Pandemic (March-May 2020) but decided to gradually open up schools as off September 2020. In a close collaboration with the UNICEF country office with a US$70,000 grant from the GPE, a rapid household survey was conducted to evaluate education sector interventions and requirements (GPE, 2020). The survey describes the challenges for distance-learning programmes vividly (United Nations Uzbekistan, 2020), as many households have no access to either digital devices, stable internet connection, or the digital literacy how to use digital platforms (Figure 3). Based on other studies by the UNICEF office, television was identified as the key media for delivering distance learning material (ibid.). Consequently, the Ministry of Public Education has taken a hybrid approach by offering video lessons on TV, in Telegram messenger application, and on the Ministry's digital platform, accompanied by feedback and assessment by teachers entrusted with supplementing lessons (ibid.).

**Table 1. Access to online / digital resources and knowledge of using online resources**

| | Rural | Urban | TOTAL |
|---|---|---|---|
| Households with television (ordinary or LCD) | 98.4% | 97.7% | 98.1% |
| Households with smart phone | 70.9% | 83.5% | 77.2% |
| Households with computer (desktop / laptop) | 19.4% | 39.3% | 29.3% |
| Households with Internet (mobile / landline) | 30.2% | 50.5% | 40.2% |
| School children with computer literacy | 37.4% | 47.6% | 42.2% |
| School children with digital literacy | 29.6% | 46.4% | 37.6% |

**Figure 3. Access to online/digital resources and knowledge of using online resources (United Nations Uzbekistan, 2020)**

## 3. Next Steps

In the next steps of this research project, the resilience attributes of a resilient ICT4D would be applied to education responses described above. The resilience perspective is a helpful approach since it moves beyond a single focus on challenges and harms and identifies possible chances and opportunities in strengthening education systems in lower- and lower-middle-income countries using ICTs. Distance learning programmes offer resilient responses in countries with frequent school disruption, such as Nigeria through infrastructural development, the availability of resources in digital format, and innovative assessment methods (Oladip, Oyedele, & Fawale, 2020). Other research on open and distance learning in developing world (Perraton, 2006) shows the potential of e-learning in assisting post-crisis countries in rebuilding their education systems (see, for example, Rhema & Miliszewska, 2012, for a discussion on Libya), or in crisis-prone countries such as Lebanon (Baytiyeh, 2019) or in war zones such as Syria, where children experience insecurity, instability, lack of resources, and lack of adult supervision (Almasri, Tahat, Skaf, & Masri, 2019). Open and distance learning programmes could also be regarded as a response to the global crisis in teacher education and training (Banks, Moon, & Wolfenden, 2009).

Any emergency always effects the target population unequally. Researchers have already discussed the setbacks of crisis, especially on educating minority groups (see, for example, Ingubu, 2010, for a discussion on Sub-Saharan Africa). However, a thorough response analysis to COVID-19 as a global emergency situation can provide governments, international organisations, and regional institutions with tools and lessons for future planning and policy making. For instance, the current crisis has once more underlined the issue of girls' education in ECO region. Recent studies have demonstrated how the lack of private space at home affects female college students in Palestine and facilitates returning to traditional gender roles and family obligations; therefore, putting an obstacle on their ability to study at home (Meler, 2021). As already discussed in the country profiles in Pakistan and Afghanistan, the effective educational platforms have increased girls' attendance but have also increased concerns that digital means would pave the way for prohibiting girls' school attendance in light of digital home schooling. Consequently, this project aims to highlight the complexities of crisis response in education sector especially in countries with existing vulnerabilities in educational systems.

## REFERENCES AND CITATIONS